# A statistical analysis of acoustic emission signals for tool condition monitoring (TCM)

G. Pontuale, F.A. Farrelly, A. Petri and L. Pitolli

*Consiglio Nazionale delle Ricerche, Istituto di Acustica "O.M.Corbino", 00133 Rome, Italy*
giorgio@idac.rm.cnr.it

**Abstract:** The statistical properties of acoustic emission signals for tool condition monitoring (TCM) applications in mechanical lathe machining are analyzed in this paper. Time series data and root mean square (RMS) values at various tool wear levels are shown to exhibit features that can be put into relation with ageing in both cases. In particular, the histograms of raw data show power-law distributions above a cross-over value, in which newer cutting tools exhibit more numerous larger events compared with more worn-out ones. For practical purposes, statistics based on RMS values are more feasible, and the analysis of these also reveals discriminating age-related features. The assumption that experimental RMS histograms follow a Beta (β) distribution has also been tested. The residuals of the modeling β functions indicate that the search for a more appropriate fitting function for the experimental distribution is desirable.
©2002 Acoustical Society of America

## 1. Introduction

The challenge of extensively using TCM devices and techniques in industrial manufacturing processes has become ever more important. In this field, a continuous improvement in terms of reliability and practicality of sensor fusion systems and algorithms is the actual challenge [Govekar *et al*., 2000]. In fact, advanced sensor design, coupled with signal processing technologies, enables one to have more refined information regarding process conditions, so that optimized control leads to product quality improvements, reduced non-productive times, and thus, considerable cost savings.

An excellent review of the state-of-the-art, technological challenges and future developments of these systems is described by [Byrne *et al*.,1995]. This paper describes in great detail the physical parameters to be analyzed for industrial control applications, together with their appropriate sensory systems. Among these, acoustic emission (AE) signal analysis has been demonstrated to be one of the most efficient TCM techniques that can be applied to machining processes control, as the impressive amount of literature on this subject shows. Using this framework, our paper tackles the problem of gaining greater insight into the fundamental statistical properties of AE signals generated during turning machining processes, so that an appropriate implementation of AE sensor-based devices will lead to efficient TCM systems.

Results of the effects of tool wear on the signal's statistical characteristics have been studied, both on their time series and RMS (root mean squared) values.

## 2. Operational conditions and experimental set-up

Measurements have been performed of machining stainless steel (AISI 303) bars on a SAG14 GRAZIANO lathe machine. Cutting speeds range from 0.5 to 1 m/s, while feed rates and cutting depths are kept constant at 0.0195 mm/turn and 2 mm respectively. These low-fatigue machining conditions have been chosen for the purposes of the present paper. In all measurements, cutting tool inserts are "IMPERO" PCLNR with 2020/12 type tungsten carbide; the acquisitions have been performed on inserts with various degrees of wear. Specifically, inserts have been grouped into three different wear categories: new ones, those estimated to be half-way through their life-cycle (50%), and those completely worn through (100%). In the new and 100% worn cases, 8 cutting edges were analyzed per wear level, whereas only 4 edges were utilized in the 50% case. For each edge, one acquisition run was conducted, collecting 15 banks of 40.960 AE time series points corresponding to 16.38 ms, for a total of 614.400 points for run. Hence, a total of 12.288.000 (4.9152s) AE time series points were collected over the 20 runs. To achieve this, a custom-built AE sensor was used, made of a small rectangular-shaped piezoelectric ceramic (PZT-5), 5.0x1.5x0.7 mm in size, housed inside a little cavity bored into the cutting tool holder to protect it from chip damages and liquid coolant effects, and placed about two centimeters from the AE signal sources. In fact, the propagation of AE signals in the investigated range is characterized by significant attenuation. Thus, to achieve a good signal-to-noise ratio, the sensor should be placed as close as possible to the machining point where the AE signal is generated [Jemielniak, 2000]. As an added benefit, reduction of the signal distortion due to the number of interfaces and mechanical resonance is also achieved by avoiding a long measurement chain. An electrically conductive adhesive is used to bond the ceramic to the internal face of the cavity. The actual experimental set-up is roughly sketched in Fig. 1.

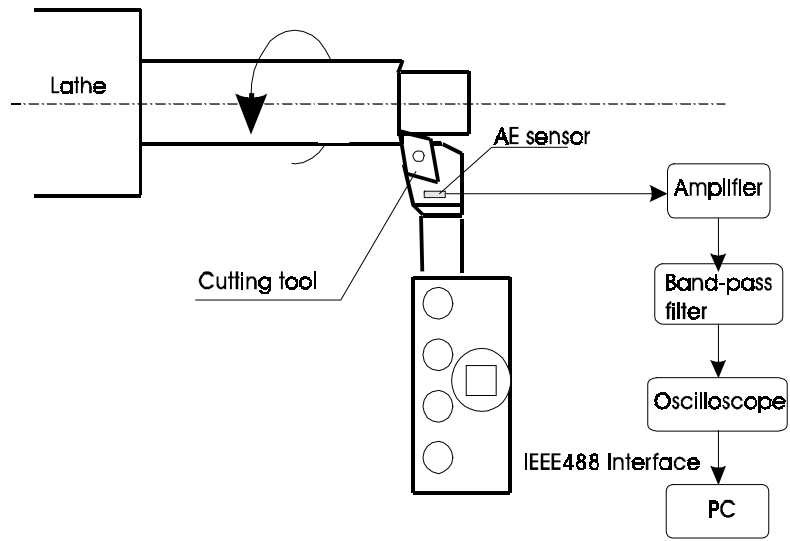

Fig. 1. Schematic diagram of experimental set-up.

The signal detected by the transducer is amplified and filtered in the 200 kHz-1000 kHz range. The filtered signal is then captured by a Tektronix digital oscilloscope (TDS420) using a 2.5 MHz sampling rate and finally stored in a Personal Computer through an IEEE488 interface. Blank measurements performed just prior to machining indicate no significant electrical noise. The data have been analyzed both directly in their time series form and through root mean squared (RMS) values.

### 3. Experimental results

*3.1 Time series analysis*

The histograms of the absolute value of time series amplitudes taken from measurements performed using inserts in three stages of wear are portrayed in Fig. 2.

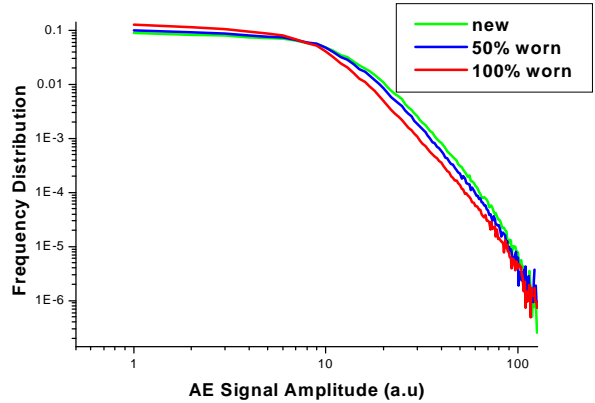

Fig.2. Histograms of the absolute value of time series amplitudes for the three wear levels.

These frequency distributions are normalized to their area and grouped into 126 classes. The curves exhibit a power-law behavior:

$$p(n) = Ax^{-\alpha} + B$$

above a cross-over value. The value of the slope (nearly $\alpha=-3.7$) is similar for all three stages of wear, and the corresponding exponent for the energy is $\alpha=-2.35$, because it can be derived from the amplitude exponent. In this range, data from measurements using tools with greater wear show a smaller frequency count for a given value in amplitude. This leads to the conclusion that in this set of trials, the newer tools are the most active ones in terms of acoustic emission. It is interesting to note that this power-law behavior has been observed in many studies on acoustic emission signals, e.g,. those associated with the formation of micro-fractures [Petri, 1996]. Power-law characteristics are associated with scale invariant properties underlying the physical phenomena under study, and in some cases, this has been explained by self-organized criticality (SOC) [Bak *et al*., 1997] models. These promising results should stimulate future investigations of these AE power-law features correlated with tool wear.

*3.2 RMS analysis*

The analysis of the RMS has been conducted calculating values on the basis of 100 points, corresponding to 40 µs. This makes this choice effective in making RMS signal sensitive to the different contributions from burst and continuous events. A substantial effort in the past has been dedicated to analyzing the relationship between RMS and tool wear level in various experimental situations, e.g., see [Kannatey-Asibu and Dornfeld, 1982] for high-speed steel tools and [Jemielniak and Otman, 1998] for identifying catastrophic tool failure (CTF) conditions in carbide inserts. To obtain a deeper understanding of the RMS properties as a function of ageing, we have analyzed their frequency distribution by grouping the values into 60 classes. For each wear level, the average histograms with their error bars are shown on the left side of Fig.3. Each bin value is obtained by dividing it by the largest of the entire sample set, and the curves are normalized to the unit area. The curves show a noticeable shift toward lower levels of the modal value of the frequency distribution for increasing levels of wear, and a change in the skewness tending toward values compatible with a symmetrical shape. To test the difference among these graphs, we have performed a T-test analysis regarding the sample means. The null hypothesis of equal means can be rejected with a confidence level of 95%. This approach seems to be effective in showing discriminating tool wear features and could be used as the basis for implementing algorithms for TCM systems.

In the literature, borrowing from a technique used in the description of surface roughness [Whitehouse, 1978], various attempts have been made to determine tool condition relying on the hypothesis that a Beta distribution $f(x)$ [Kannatey-Asibu and Dornfeld, 1982] [Jemielniak and Otman, 1998], properly describes the probability density function *pdf* of the RMS values:

$$f(x) = \frac{x^{r-1}(1-x)^{s-1}}{\beta(r,s)} \quad (1)$$

where $\beta$ is the complete Beta function:

$$\beta(r,s) = \int_0^1 x^{r-1}(1-x)^{s-1} dx. \quad (2)$$

With this assumption, it is possible to characterize the moments of the distribution in terms of the two parameters $r$ and $s$, and vice-versa. In particular, as far as mean $\mu$ and variance $\sigma^2$ are concerned, we have:

$$r = \frac{\mu}{\sigma^2}(\mu - \mu^2 - \sigma^2)$$
$$s = \frac{1-\mu}{\sigma^2}(\mu - \mu^2 - \sigma^2) \quad (3)$$

Thus, values for $r$ and $s$ can be estimated on the basis of the mean and variance of the data set. Past studies have shown that $r,s$ pairs are scattered in different ways, depending on tool conditions [Kannatey-Asibu and Dornfeld, 1982]. One shortcoming of this method is that no estimate of the errors on the $r$ and $s$ parameters is directly available; this is particularly serious because real-life signals often contain outliers that can bring a noticeable shift in the actual values of both mean and variance. One possibility is to use more robust estimators (e.g., median instead of mean), although this still does not give an error estimate for the calculated parameters. A further choice is to perform a nonlinear best-fit on the data set using the function given in Eq.(1).

Fitting the function given in Eq.(1) requires a nontrivial effort, because this function cannot be linearized, so that the fitting process requires initial seed values; these might be obtained using the $r,s$ estimation technique described above. Additionally, fully automating the fitting process requires appropriate handling of specific exceptions that can occur in nonlinear regressions (e.g., singular matrix conditions, reaching iteration limits without having satisfied tolerance criteria, etc.). In this paper, no attempt to automate this process has been made.

In the right side of Fig.3, the best-fit of the frequency distributions of the same data sets as in the left part are shown. From these graphs, we see that although there is a good matching between the fitting function and the data sets in the neighborhood of the peaks, some discrepancies are visible in the residual for RMS bin values just above the peak where the curves level off. This indicates that in this range, the data sets are richer in events than Eq.(1) indicates. This suggests that a better empirical fitting-function may exist. In Fig.4 $r,s$ estimates from Eq.(2) are compared to the ones obtained by the best-fitting process. It is evident that the two groups differ greatly and that these discrepancies are not compatible considering the error estimates given on the fitted parameters. Furthermore, the scattering

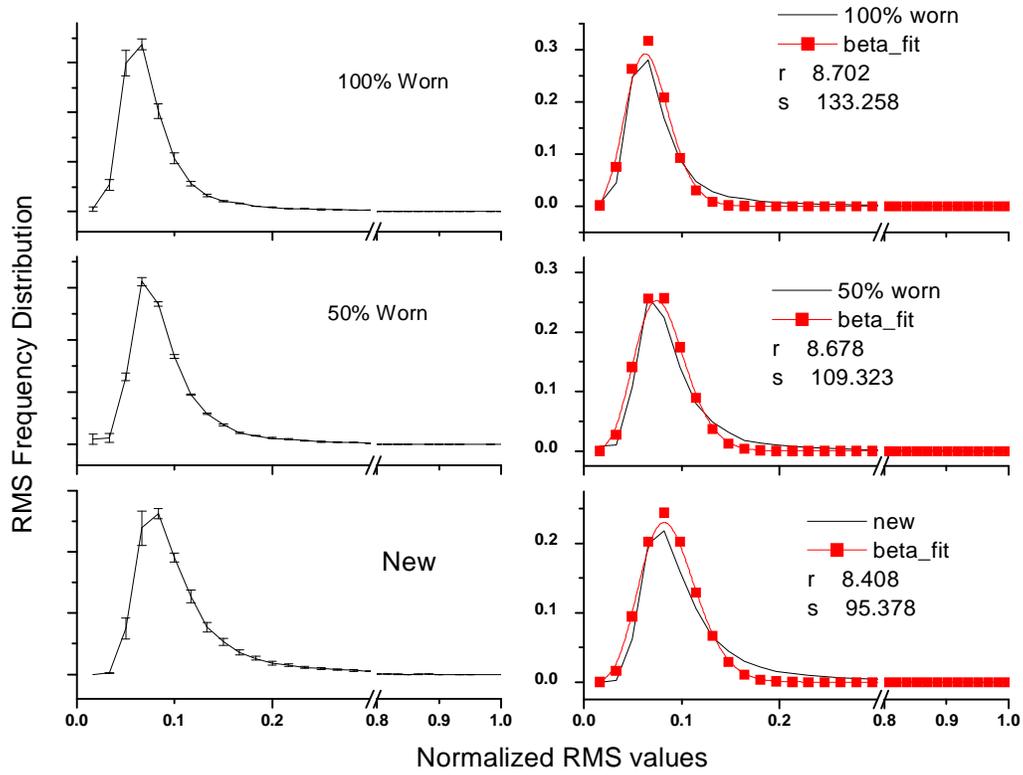

Fig3. (*left*):, Average RMS histograms with associated standard error shown for each wear level; (*right*): Same frequency distributions shown together with their β-function best-fit.

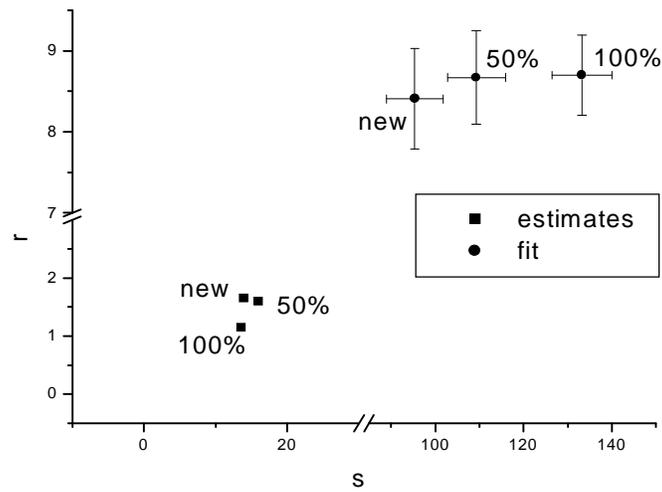

Fig.4. *r,s* estimates from Eq.(2) compared to the ones obtained by the best-fitting process

pattern of these two groups are entirely different; whereas both the best-fitted *r,s* parameters tend to increase with wear, the estimated ones show an essentially opposite behavior. One possible explanation for this difference is that while the best-fit process minimizes mean-square differences between the fitting function and the frequency distribution (so that heavily populated bins are weighted more), the estimate method relies on $\mu$ and $\sigma^2$. Variance, in particular, is highly sensitive to outliers. Values far from the mean weigh heavily on its determination.

### 4. Discussion and conclusions

Various ways of analyzing the statistical properties of AE signals in a TCM application have been illustrated. In particular, we have seen that it is possible to observe tool wear level related features both in AE time series and their RMS values. Particularly interesting are the statistical properties of the former, in which power law characteristics have been identified through the use of log-log histograms. This behavior has already been observed in the properties of acoustic emission signals in numerous other fields.

The frequency distributions of the RMS values have also been studied as a function of wear, showing that even in this case it is possible to identify discriminating features. Moreover, a beta function model for describing the RMS's *pdf* has been tested. We have compared the parameters *r,s* of the beta distribution obtained by a best-fit procedure to their estimates computed directly. The residuals in the best-fitted function indicate that a more appropriate fitting model should be sought, and that the comparison of the parameters obtained in this way are entirely different from those estimated through the statistical moments of the data. For these reasons, future efforts will be dedicated in investigating the theoretical form of the *pdf* of RMS values related to power-law distributed time series.

**Acknowledgment**
The authors gratefully acknowledge the valuable support of IMPERO S.p.A. Furthermore, we wish to express gratitude to our colleague and friend Mario Acciarini, who prepared the sensors and has recently died.